\documentclass{PoS}
\newcommand{\lat}{\mathrm{lat}}
\newcommand{\fslash}[1]{\slash\!\!\!\!\!{#1}}   
\newcommand{\Dlr}{\stackrel{\leftrightarrow}{D}}
\newcommand{\dlangle}{\langle\!\langle} 
\newcommand{\drangle}{\rangle\!\rangle}
\newcommand{\imag}{\mbox{i}}  

\title{Nucleon structure in the chiral regime with domain wall
fermions on an improved staggered sea}

\ShortTitle{Nucleon structure in the chiral regime with domain wall fermions \hspace{4pc} J.W.~Negele and D.B.~Renner}

\author{LHPC Collaboration: R.G.~Edwards$^a$, G.~Fleming$^b$, Ph.~H\"agler$^c$,
J.W.~Negele\thanks{Speakers.} \,$^d$,
K.~Orginos$^{a,e}$, A.V.~Pochinsky$^d$, D.B.~Renner$^{*f}$, D.G.~Richards$^a$, W.~Schroers$^g$\\
\llap{$^a$}Thomas Jefferson National Accelerator Facility, Newport News, VA 23606, USA\\
\llap{$^b$}Sloane Physics Laboratory, Yale University, New Haven, CT 06520, USA\\
\llap{$^c$}Institut f\"ur Theoretische Physik, TU M\"unchen, D-85747 Garching, Germany\\
\llap{$^d$}Center\hspace{-0.0725pc} for\hspace{-0.0725pc} Theoretical\hspace{-0.0725pc} Physics,\hspace{-0.0725pc} Massachusetts\hspace{-0.0725pc} Institute\hspace{-0.0725pc} of\hspace{-0.0725pc} Technology,\hspace{-0.0725pc} Cambridge,\hspace{-0.0725pc} MA\hspace{-0.0725pc} 02139,\hspace{-0.0725pc} USA\\
\llap{$^e$}Department of Physics, College of William and Mary, Williamsburg VA 23187, USA\\
\llap{$^f$}University of Arizona, Department of Physics, 1118 E 4th St, Tucson AZ 85721, USA\\
\llap{$^g$}John von Neumann-Institut f\"ur Computing NIC/DESY, D-15738 Zeuthen, Germany}

\abstract{Moments of unpolarized, helicity, and transversity
distributions, electromagnetic form factors, and generalized form
factors of the nucleon are presented from a preliminary analysis of
lattice results using pion masses down to 359 MeV.  The twist two
matrix elements are calculated using a mixed action of domain wall
valence quarks and asqtad staggered sea quarks and are renormalized
perturbatively.  Several observables are extrapolated to the physical
limit using chiral perturbation theory. Results are compared with
experimental moments of quark distributions and electromagnetic form
factors and phenomenologically determined generalized form factors,
and the implications on the transverse structure and spin content of
the nucleon are discussed.}

\FullConference{XXIVth International Symposium on Lattice Field
Theory\\July 23-28, 2006\\ Tucson, Arizona, USA}

\begin{document}

\section{Introduction}

Our use of lattice field theory to calculate the structure of the
nucleon from first principles has two complementary objectives.  One
goal is to achieve quantitative agreement with experimental
observables such as nucleon form factors and parton distributions, to
confirm the quantitative precision of our solution of QCD and to
establish the credibility to make predictions and guide future
experiments.  However, merely producing a black box that only
reproduces experiment would be unfulfilling. Hence, our second goal is
to obtain insight into how QCD works, revealing, for example, the
origin of the nucleon spin, physical mechanisms, such as instantons,
responsible for essential features of hadron structure, and the
dependence on parameters of QCD like $N_C$, $N_f$, $m_q$, and the
gauge group.  A crucial issue in making quantitative contact with
experiment is calculating sufficiently far into the chiral regime that
reliable chiral extrapolations to the physical pion mass are possible.
Hence, in this work we have utilized the extensive set of
configurations with dynamical improved staggered quarks generated by
the MILC collaboration~\cite{Bernard:2001av} to explore nucleon structure in the chiral
regime.

\section{Mixed Action Lattice Calculation}

As explained in Ref.~\cite{Renner:2004ck}, we utilize a hybrid action
combining domain wall valence fermions with improved staggered sea
quarks.  Improved staggered sea quarks offer the advantage that due to
the relative economy of the algorithm, lattices with large volumes,
small pion masses, and several lattice spacings are publicly available
from the MILC collaboration.  Although the fourth root of the fermion
determinant remains controversial, current evidence suggests it is
manageable~\cite{Bernard:2006ee,Sharpe:2006re}.  Renormalization group
arguments indicate that the coefficient of the
nonlocal term approaches zero in the continuum limit~\cite{Shamir:2006nj}, partially
quenched staggered chiral perturbation theory accounts well for the
artificial properties at finite lattice spacing~\cite{Bernard:2006zw}, and the action has
the advantage of being improved to ${\cal{O}}(a^2)$. Domain wall
valence quarks offer equally compelling advantages to justify
investing resources in calculating hadron observables on staggered
configurations that are roughly comparable to the resources required
to generate the configurations themselves.  Domain wall fermions
prevent mixing of quark observables by chiral symmetry, are accurate
to ${\cal{O}}(a^2)$, and possess a conserved five dimensional axial
current that facilitates calculation of renormalization factors. In
addition, hybrid action (often referred to as mixed action) chiral
perturbation theory results are available for many observables, and by
virtue of an exact lattice chiral symmetry, one loop results have the
simple chiral behavior observed in the continuum.  The parameters of
the configurations used in this work are shown in Table~[\ref{table}],
and details can be found in Refs.~\cite{Renner:2004ck,Edwards:2005kw}.
\begin{table}[htb]
\begin{center}
\begin{minipage}{18pc}
\begin{center}
\begin{tabular}{|l|l|l|l|l|} \hline
$a m_{u/d}^{\mathrm{asqtad}}$ & $L/a$ & $L$ & $m_\pi^{\mathrm{DWF}}$ & \# \\ \hline
 & & $\mathrm{fm}$ & $\mathrm{MeV}$ & \\ \hline
$0.05$ & $20$ & $2.52$ & $761$ & $425$ \\ \hline
$0.04$ & $"$  & $"$    & $693$ & $350$ \\ \hline
$0.03$ & $"$  & $"$    & $594$ & $564$ \\ \hline
$0.02$ & $"$  & $"$    & $498$ & $486$ \\ \hline
$0.01$ & $"$  & $"$    & $354$ & $656$ \\ \hline
$0.01$ & $28$ & $3.53$ & $353$ & $270$ \\ \hline
\end{tabular}
\end{center}
\caption{\label{table}Lattice parameters used in this work.}
\end{minipage}
\end{center}
\end{table}

\section{Moments of Parton Distributions}

Parton distributions measure forward matrix elements of the gauge
invariant light cone operators
\begin{equation}
 {\cal O}_\Gamma(x) =\int \!\frac{d \lambda}{4 \pi} e^{i \lambda x} \overline
  q (-\lambda n/2)
  \Gamma \,{\cal P} e^{-ig \int_{\lambda / 2}^{-\lambda / 2} d \alpha \, n
    \cdot A(\alpha n)}\!
  q(\lambda n/2),
\label{LCop}
\end{equation}
where $x$ is a momentum fraction, $n$ is a light cone vector and
$\Gamma = \fslash{n}$ or $\Gamma =\fslash{n} \gamma_5$.  Using the
operator product expansion, the operators in Eq.~[\ref{LCop}] yield towers
of symmetrized, traceless local operators that can be evaluated on a
Euclidean lattice
\begin{equation}
 {\cal O }_{[\gamma_5]}^{\{\mu_1\ldots\mu_n\}}
  =\overline q  \gamma^{\{\mu_1} [\gamma_5] i\Dlr{}^{\!\mu _{2}}\!\cdots i\Dlr{}^{\!\mu _{n}\}} q\,,
\label{LocalOp} 
\end{equation}
where $[\gamma_5]$ denotes the possible inclusion of $\gamma_5$, the
curly brackets represent symmetrization over the indices $\mu_i$ and
subtraction of traces, and $\Dlr=1/2
(\stackrel{\rightarrow}{D}-\stackrel{\leftarrow}{D})$. A related
operator for transversity distributions is
\begin{equation}
 {\cal O }_{\sigma}^{\mu \{\mu_1\ldots\mu_n\}}
  =\overline q \gamma_5 \sigma^{\mu \{\mu_1}  i\Dlr{}^{\!\mu _{2}}\!\cdots i\Dlr{}^{\!\mu _{n}\}} q\,.
  \label{LocalOpt}
\end{equation}
Using the notation and normalization of Ref.~\cite{Dolgov:2002zm}, the
forward matrix elements $\langle P,S| {\cal O
}^{\{\mu_1\ldots\mu_{n+1}\}} |P,S\rangle$ yield moments of the
unpolarized quark distribution:
\begin{equation}
\langle x^n \rangle_q  =  \int_0^1\!\!dx\, x^n [q(x) + (-1)^{n+1}\overline{q}(x)],
\end{equation}
the forward matrix elements $\langle P,S| {\cal O
}_{\gamma_5}^{\{\mu_1\ldots\mu_{n+1}\}} |P,S \rangle $ yield moments
of the helicity distribution:
\begin{equation}
\langle x^n \rangle_{\Delta q}  =  \int_0^1\!\!dx\, x^n [\Delta q(x)+ (-1)^{n}\Delta \overline{q}(x)], 
\end{equation}
and the forward matrix elements $\langle P,S| {\cal O }_{\sigma}^{\mu
\{\mu_1\ldots\mu_{n+1}\}} |P,S \rangle $ yield moments of the
transversity distribution:
\begin{equation}
\langle x^n \rangle_{\delta q}  =  \int_0^1\!\!dx\, x^n [\delta q(x)+ (-1)^{n+1}\delta \overline{q}(x)].
\end{equation}
In this work, we calculate only connected diagrams, and hence
concentrate as much as possible on isovector quantities.

All quark bilinear operators in Eqs.~[\ref{LocalOp}] and
[\ref{LocalOpt}] are renormalized as follows~\cite{Bistrovic}.  The
axial current is renormalized exactly using the conserved five
dimensional axial current.  By virtue of the suppression of loop
integrals by HYP smearing, the ratio of the one-loop perturbative
renormalization factor for a general bilinear operator to the
renormalization factor for the axial current is within a few percent
of unity, suggesting adequate convergence at one-loop level. Hence the
complete renormalization factor is written as the exact axial current
renormalization factor times the ratio of the perturbative
renormalization factor for the desired operator divided by the
perturbative renormalization factor for the axial current.

\subsection{Chiral Perturbation Theory}

Ideally, we would like to perform high statistics calculations at pion
masses below 350 MeV and extrapolate them in pion mass and volume
using a chiral perturbation theory expansion of sufficiently high
order to provide a quantitatively controlled approximation. In
practice, our most convincing chiral extrapolation has been for $g_A$
using the finite volume results including $\Delta$ intermediate states
of Ref.~\cite{Beane:2004rf}, where the fit involving 6 low energy
parameters yielded an excellent fit up to the order of a 700 MeV pion
mass and agreed with experiment with 6.8\%
errors~\cite{Edwards:2005ym}.  Similar extrapolations of $g_A$ have
been performed by other groups~\cite{Khan:2006de}.  This success for
$g_A$ is particularly relevant to the subsequent discussion of nucleon
spin, because it involves the same operators $\langle 1 \rangle_q $ as
$\Delta \Sigma$. We note that because the nucleon and $\Delta$ should
be included together at large $N_c$~\cite{Dashen:1993jt} and indeed
show large cancellations in the axial charge, we prefer to include the
$\Delta$ as an explicit degree of freedom in the analysis.

An unresolved puzzle in calculating moments of structure functions is
the relatively flat behavior of the momentum fraction $\langle x
\rangle$ at a constant value substantially higher than experiment~\cite{Detmold:2001jb,Orginos:2005uy}.
Hence, it is particularly interesting to ask whether a chiral
perturbation theory fit determined without knowledge of the
experimental result is in fact statistically consistent with
experiment.  Since there is presently insufficient data to perform a
full analysis including the $\Delta$, here we present a simple
self-consistent improved one-loop analysis using only nucleon degrees
of freedom that appears to work very well in our regime.

The details will be presented in a future publication~\cite{Renner},
but the basic idea is as follows.  We begin with the one loop
expression at scale $\mu$~\cite{Arndt:2001ye,Chen:2001eg}.
\begin{equation}
\langle x^n \rangle_{u-d} = a_n \left( 1 - \frac{(3 g^2_{A,0} + 1)}{(4\pi f_{\pi,\scriptscriptstyle{0}})^2} m_\pi^2 \ln \left( \frac{m_\pi^2}{\mu^2} \right) \right) + b^\prime_n(\mu) m_\pi^2
\end{equation}
in which we explicitly note that $g_{A,0}$ and
$f_{\pi,\scriptscriptstyle{0}}$ are $g_A$ and $f_\pi$ in the chiral
limit.  We are free to choose the scale $\mu$ to be $f_\pi$.
Additionally we replace $g_{A,0}$ and $f_{\pi,\scriptscriptstyle{0}}$
with their values at the given pion mass $g_{A,m_\pi}$ and
$f_{\pi,m_\pi}$, so that the result may be rewritten as
\begin{equation}
\langle x^n \rangle_{u-d} = a_n \left( 1 - \frac{(3 g_{A,m_\pi}^2 + 1)}{(4\pi)^2} \frac{m_\pi^2}{f_{\pi,m_\pi}^2} \ln \left( \frac{m_\pi^2}{f_{\pi,m_\pi}^2} \right) \right) + b_n \frac{m_\pi^2}{f_{\pi,m_\pi}^2}.
\end{equation}
\begin{figure}[tb]
\begin{minipage}{17.75pc}
\includegraphics[width=17pc,angle=270,scale=0.8]{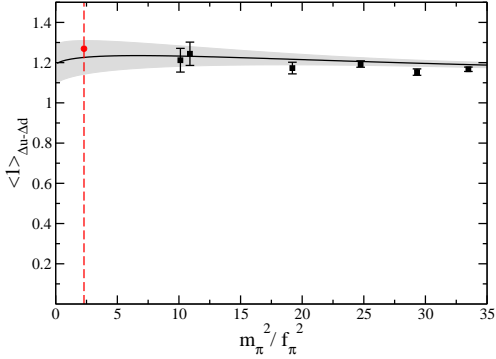}
\caption{\label{lpol_0_umd}Zeroth moment of helicity distribution.}
\end{minipage}
\hspace{0pc}
\begin{minipage}{17.75pc}
\includegraphics[width=17pc,angle=270,scale=0.8]{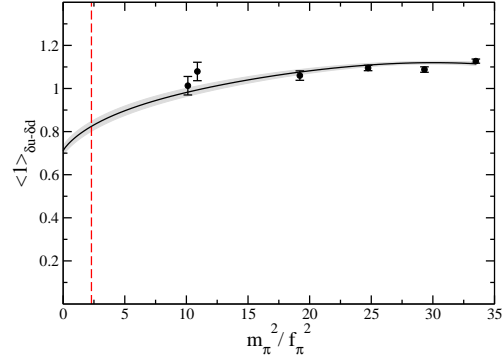}
\caption{\label{tpol_0_umd}Zeroth moment of transversity distribution.}
\end{minipage}
\end{figure}
\begin{figure}[tb]
\begin{minipage}{17.75pc}
\includegraphics[width=17pc,angle=270,scale=0.8]{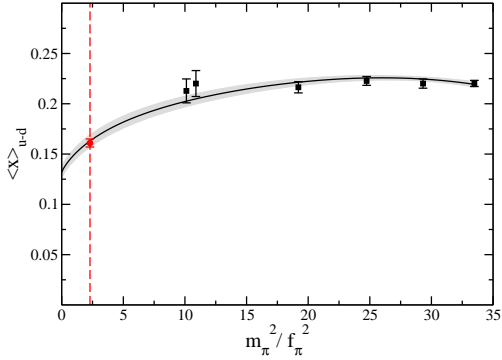}
\caption{\label{upol_1_b_umd}First moment of unpolarized distribution.}
\end{minipage}
\hspace{0pc}
\begin{minipage}{17.75pc}
\includegraphics[width=17pc,angle=270,scale=0.8]{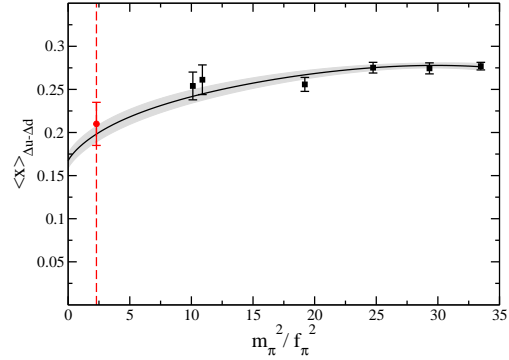}
\caption{\label{lpol_1_b_umd}First moment of helicity distribution.}
\end{minipage}
\end{figure}
If we view this as an expansion in the ratio $r =
\frac{m_\pi^2}{(4\pi)^2f^2_{\pi, m_\pi}}$, one can show that shifting
to an expansion around $g_A$ and $f_\pi$ defined at another mass only
introduces changes of ${\cal{O}}(r^2)$. Hence, to leading order, we
may write an expression in which we use the values $g_{A,lat}$,
$f_{\pi,lat}$, and $m_{\pi,lat}$ calculated on the lattice at specific
values of the quark mass. Then, the expressions for the moments of the
unpolarized, helicity, and transversity distributions are the
following:
\begin{eqnarray}
\langle x^n \rangle_{u-d} & = &
a_n \left( 1 - \frac{(3 g_{A,\lat}^2 + 1)}{(4\pi)^2} \frac{m_{\pi,\lat}^2}{f_{\pi,\lat}^2} \ln \left( \frac{m_{\pi,\lat}^2}{f_{\pi,\lat}^2} \right) \right) + b_n \frac{m_{\pi,\lat}^2}{f_{\pi,\lat}^2} \\
\langle x^n \rangle_{\Delta u-\Delta d} & = &
\Delta a_n \left( 1 - \frac{(2 g_{A,\lat}^2 + 1)}{(4\pi)^2} \frac{m_{\pi,\lat}^2}{f_{\pi,\lat}^2} \ln \left( \frac{m_{\pi,\lat}^2}{f_{\pi,\lat}^2} \right) \right) + \Delta b_n \frac{m_{\pi,\lat}^2}{f_{\pi,\lat}^2} \\
\langle x^n \rangle_{\delta u-\delta d} & = &
\delta a_n \left( 1 - \frac{(4 g_{A,\lat}^2 + 1)}{2(4\pi)^2} \frac{m_{\pi,\lat}^2}{f_{\pi,\lat}^2} \ln \left( \frac{m_{\pi,\lat}^2}{f_{\pi,\lat}^2} \right) \right) + \delta b_n \frac{m_{\pi,\lat}^2}{f_{\pi,\lat}^2}\,.
\end{eqnarray}
These results allow a least-squares two-parameter fit to the lattice
data for moments and provides an extrapolation to the physical pion
mass with a corresponding error band.  Note that the series is
substantially rearranged, by virtue that the calculated values of
$g_A$, $f_\pi$, and $m_\pi$ are used at each value of the bare quark
mass.  Although we cannot prove that this self-consistent improved
one-loop result should be accurate throughout the range of our data,
to the extent to which it is successful, we believe its success arises
from this self-consistent rearrangement.  Additionally, the use of
physical rather than chiral limit values for $f_\pi$ was first tried
in~\cite{Beane:2005rj} and has since been studied
in chiral perturbation~\cite{Chen:2005ab,O'Connell:2006sh} and applied to 
a variety of lattice
calculations~\cite{Beane:2006gj,Beane:2006kx,Beane:2006fk,Beane:2006pt,Beane:2006mx}.

\subsection{Lattice Results}

\begin{figure}[htb]
\begin{minipage}{17.75pc}
\includegraphics[width=17pc,angle=270,scale=0.8]{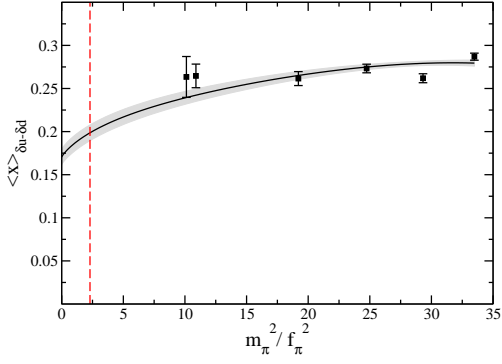}
\caption{\label{tpol_1_umd}First moment of transversity distribution.}
\end{minipage}
\hspace{0pc}
\begin{minipage}{17.75pc}
\includegraphics[width=17pc,angle=270,scale=0.8]{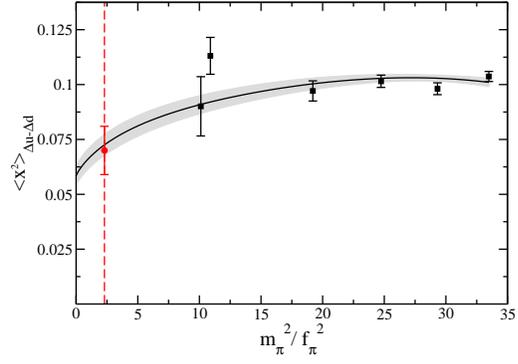}
\caption{\label{lpol_2_umd}Second moment of helicity distribution.}
\end{minipage}
\end{figure}

Here we show the results of this one-loop
analysis. Figure~[\ref{lpol_0_umd}] shows the result for $g_A$, which
is nearly as good as the complete analysis of
Ref.~\cite{Edwards:2005ym}, and yields a comparable extrapolation and
error bar.  Reassured by this result, we show analogous results for
$\langle 1 \rangle_{\delta u - \delta d}$, $\langle x \rangle_{u -
d}$, $\langle x \rangle_{\Delta u - \Delta d}$,$\langle x
\rangle_{\delta u - \delta d}$, and $\langle x^2 \rangle_{\delta u -
\delta d}$ in Figs.~[\ref{upol_1_b_umd}]-[\ref{lpol_2_umd}].

\begin{figure}[htb]
\begin{center}
\includegraphics[width=24pc]{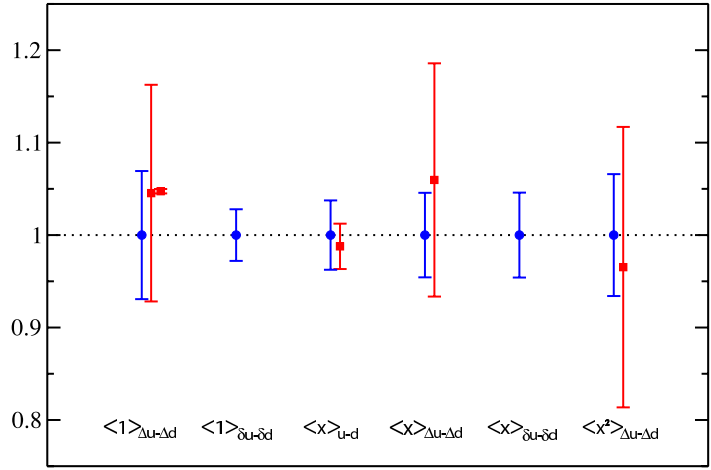}\vspace{-3pc}
\caption{\label{summary_mod}The six moments considered in this work.  Lattice results are shown in blue and experimental measurements in red, and each is normalized to the corresponding lattice result.}
\end{center}
\end{figure}

Note that in every case for which there is experimental data, this
analysis, which in no way includes the experimental result in the fit,
yields an extrapolation consistent with experiment. The results are
collected together in Fig.~[\ref{summary_mod}], where because
experimental results are not available for all cases, we have
normalized all results to the corresponding lattice result.

\section{Form Factors}

Form factors are interesting physically because at low momentum
transfer they characterize the spatial size of charge and current
distributions and at high momentum transfer they measure the ability
of the nucleon to absorb a large momentum and distribute it to all the
constituents such that the system remains in its ground state.
Although for a relativistic system, the slope of the form factor is
not precisely related to the rms radius, we will adhere to the common
usage and refer to the slope as the rms radius.  (We note in passing
that the slope of $F_1$ is in fact related to the transverse rms
radius in the infinite momentum frame.) Our primary focus here is on
the qualitative approach to experiment as the quark mass is decreased,
and to avoid uncalculated contributions of disconnected diagrams, we
consider isovector form factors.

\begin{figure}[tb]
\begin{minipage}{17.5pc}
\includegraphics[width=17pc,angle=0,scale=1.0]{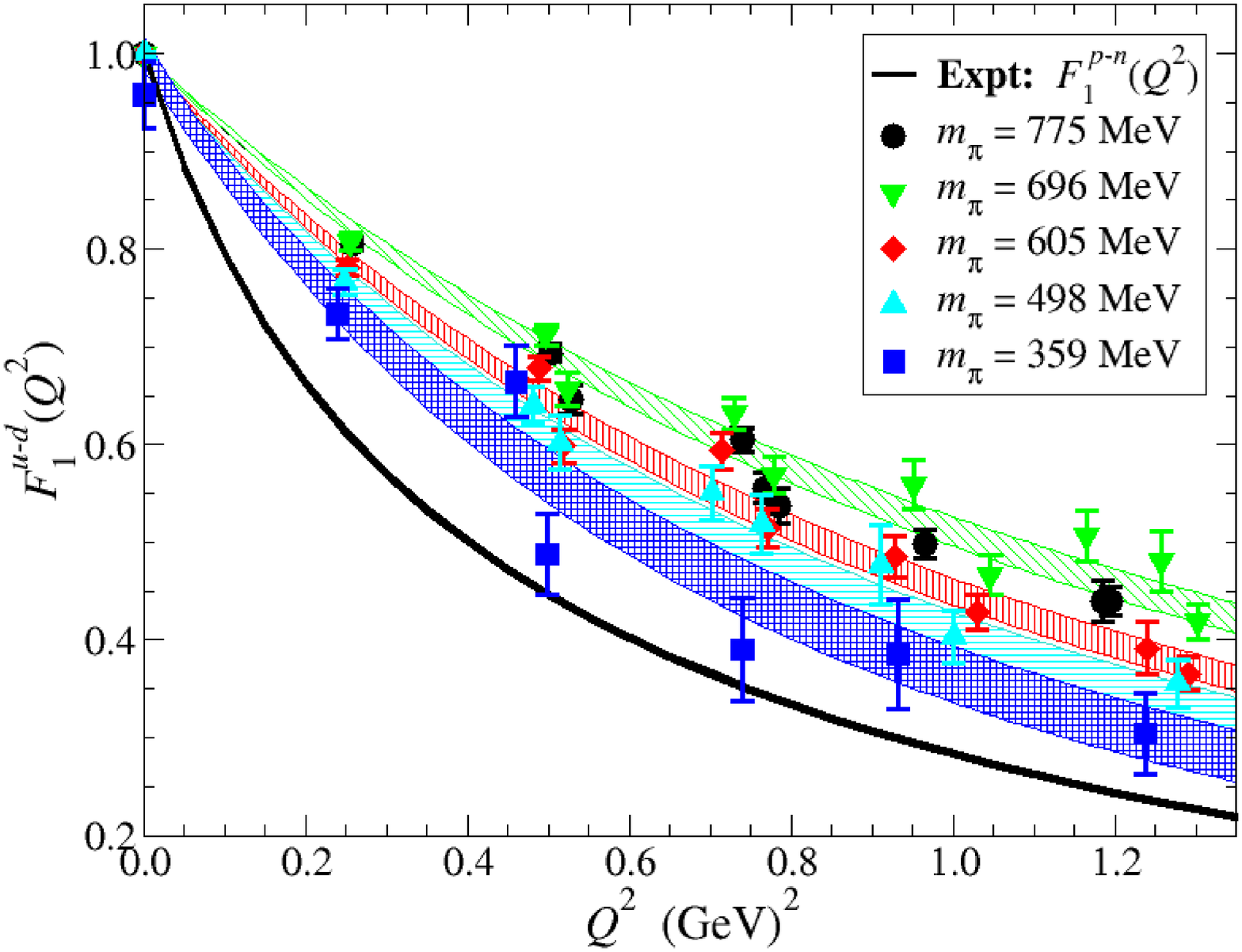}
\caption{\label{F1}F$_1$ isovector form factor at five masses compared with experiment~\cite{Kelly:2004hm}.}
\end{minipage}
\hspace{0.5pc}
\begin{minipage}{17.5pc}
\includegraphics[width=17pc,angle=0,scale=1]{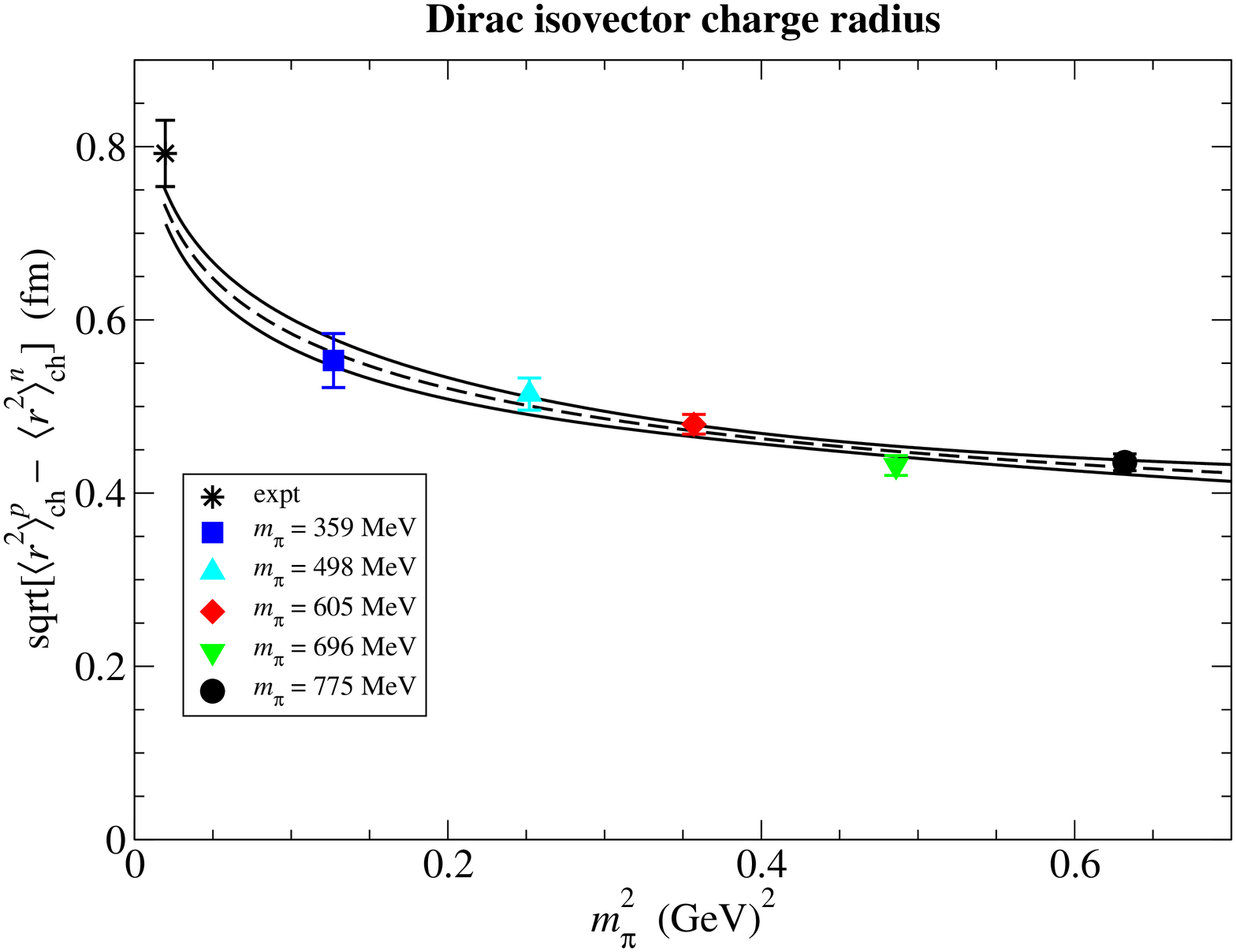}
\caption{\label{rms}Chiral extrapolation of isovector form factor slope.}
\end{minipage}
\end{figure}

The nucleon vector current form factors $F_1$ and $F_2$ are defined by
\begin{equation}
\langle p | \overline{\psi}  \gamma^\mu \psi | p' \rangle = \overline{u} (p) [F_1(q^2)\gamma^\mu  +  F_2(q^2)\frac{i \sigma^{\mu\nu} q_\nu}{2 m} ] u(p')\,.
\end{equation}
Figure~[\ref{F1}] shows the lattice data and dipole fits for $F_1$ at
five pion masses, and one observes that the lattice results
systematically approach the experimental curve as the pion mass
decreases to 359~MeV.  One can quantitatively observe how the rms
radius $\langle r^2 \rangle^{n-p}$ defined from the slope approaches
the experimental value by fitting it with the simple chiral
extrapolation formula~\cite{Dunne:2001ip}
\begin{equation}
\langle r^2 \rangle^{u-d} = a_0 -\frac{1+5g_A^2}{(4 \pi f_\pi)^2}\log\left(\frac{m^2_\pi}{m^2_\pi + \Lambda^2}\right),
\label{rms-extrap}
\end{equation}
where $\Lambda$ is a phenomenological cutoff or equivalently,
coefficient of an $m_\pi^2$ term in the expansion.  Note that, in
contrast to most chiral extrapolations which contain finite terms of
the form $m_\pi^2\log m_\pi^2$, the isovector radius diverges like
$\log m_\pi^2$, rendering the variation of the radius quite
substantial near the physical pion mass.  Figure~[\ref{rms}] shows the
results of fitting the lattice data with Eq.~[\ref{rms-extrap}], and one
notes that without having been constrained to do so, the chiral
extrapolation is consistent with experiment.
\begin{figure}[tb]
\begin{minipage}{17.5pc}
\includegraphics[width=17pc,angle=0,scale=1]{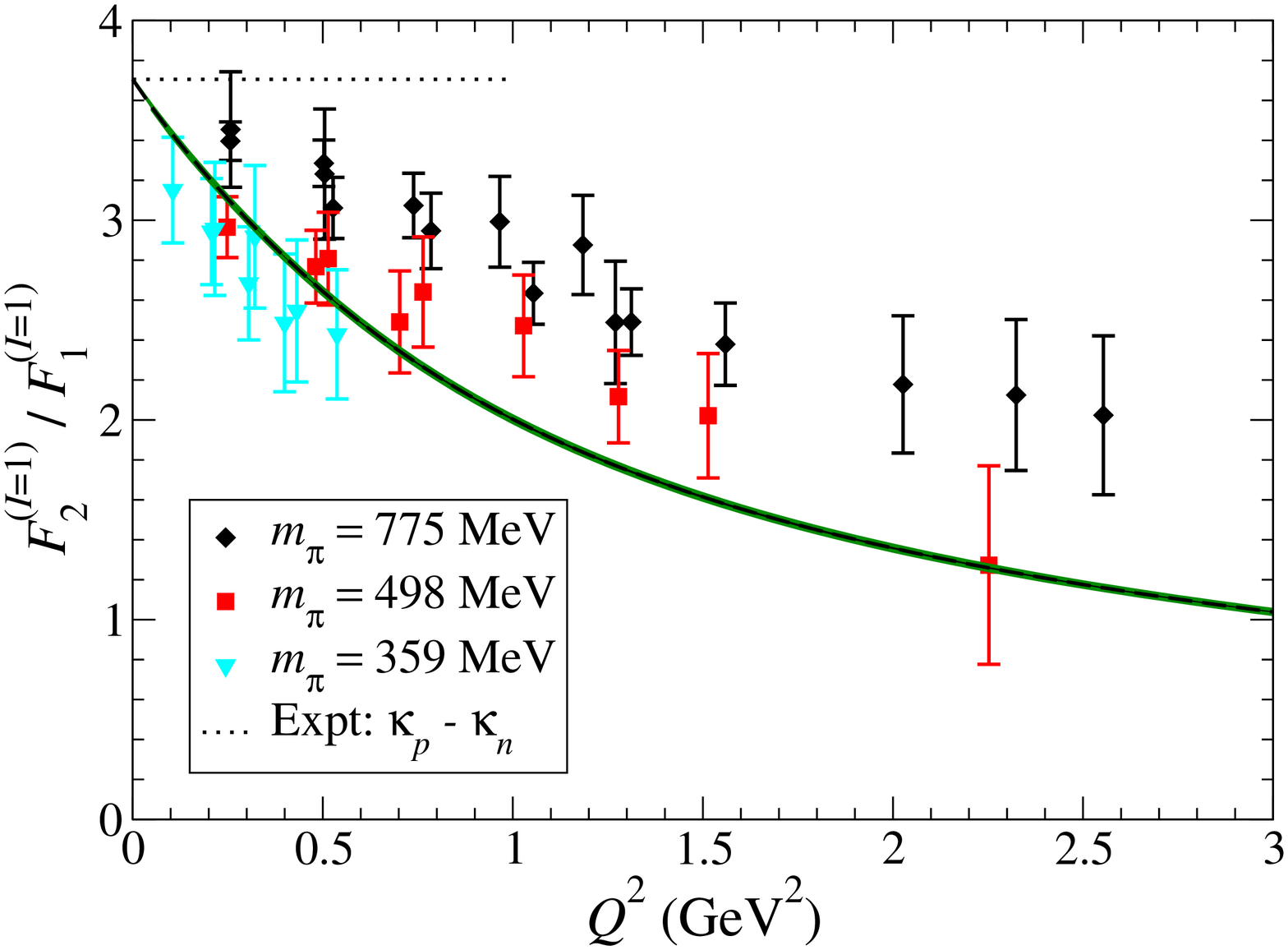}
\caption{\label{F2_F1}Isovector form factor ratio $F_2/F_1$ at three masses compared with experiment~\cite{Kelly:2004hm}.} 
\end{minipage}
\hspace{0.5pc}
\begin{minipage}{17.5pc}
\raisebox{0cm}{\includegraphics[width=17pc,angle=0,scale=1.0]{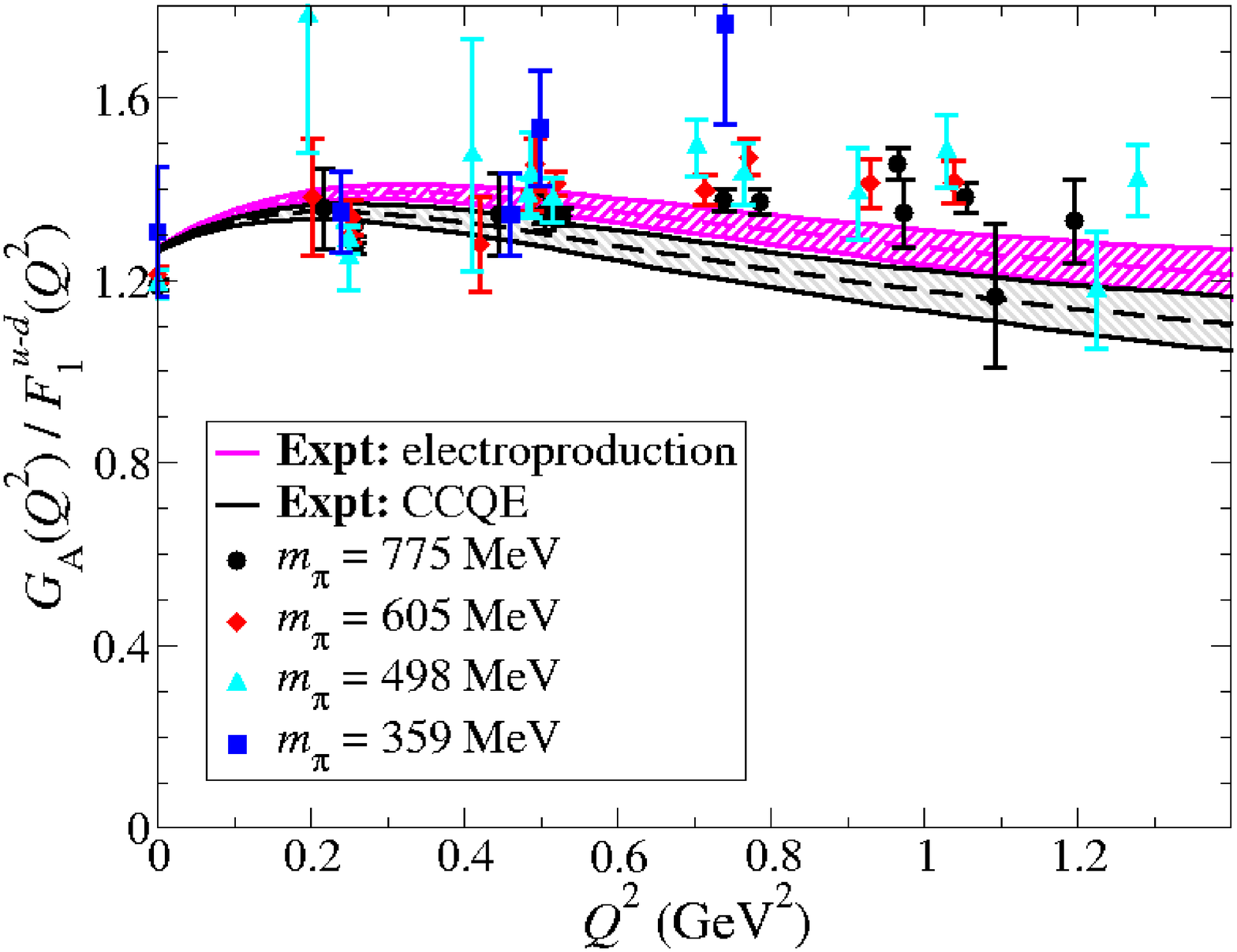}}
\caption{\label{GA_F1}Isovector form factor ratio $G_A/F_1$ at four masses compared with experiment.}
\end{minipage}
\end{figure}
\begin{figure}[tb]
\begin{minipage}{17.5pc}
\includegraphics[width=17pc,angle=0,scale=0.95]{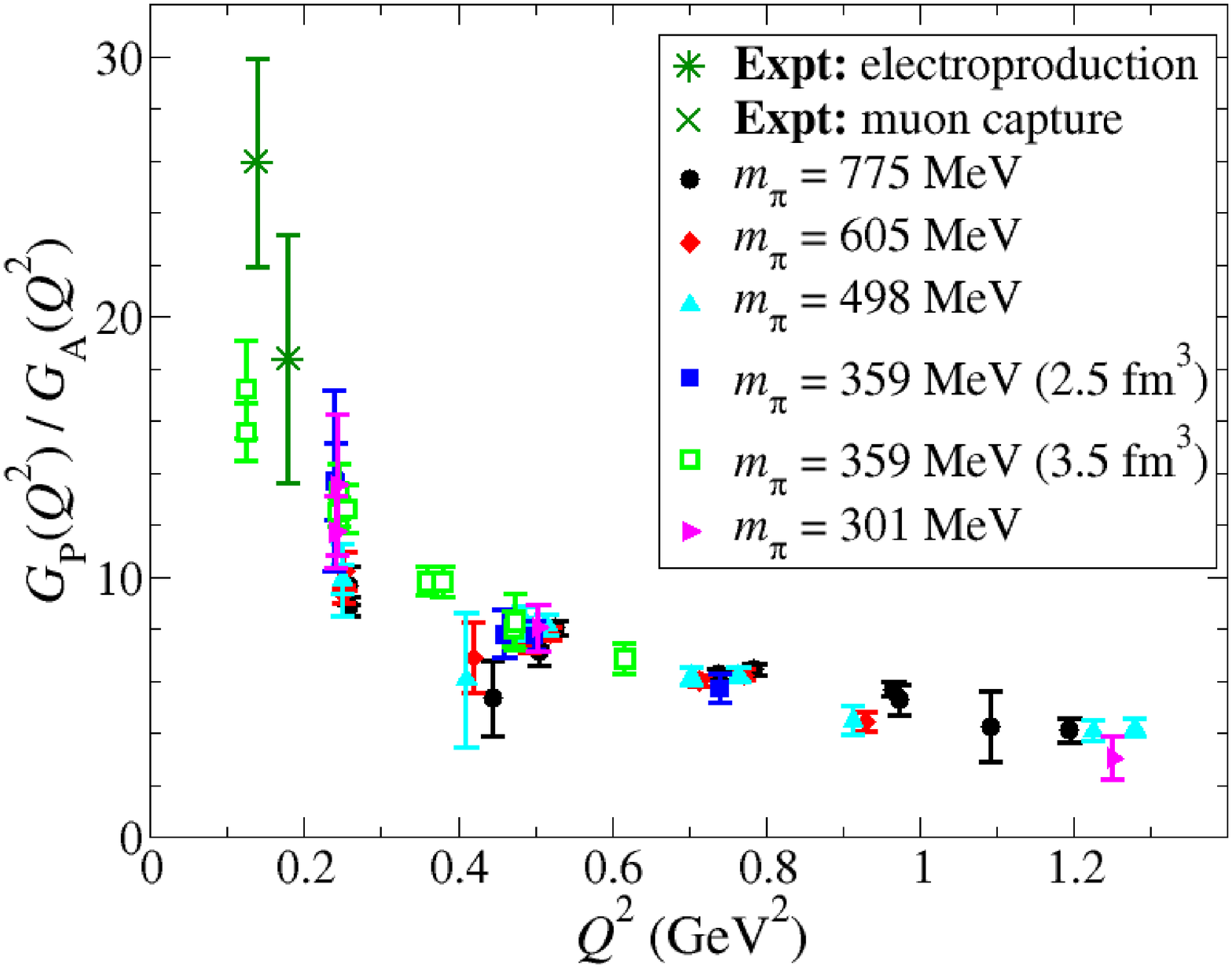}
\caption{\label{GP_GA} Isovector form factor ratio $G_P/G_A$ at six masses compared with experiment.}
\end{minipage}
\hspace{0.5pc}
\begin{minipage}{17.5pc}
\includegraphics[width=17pc,angle=-90,scale=0.79]{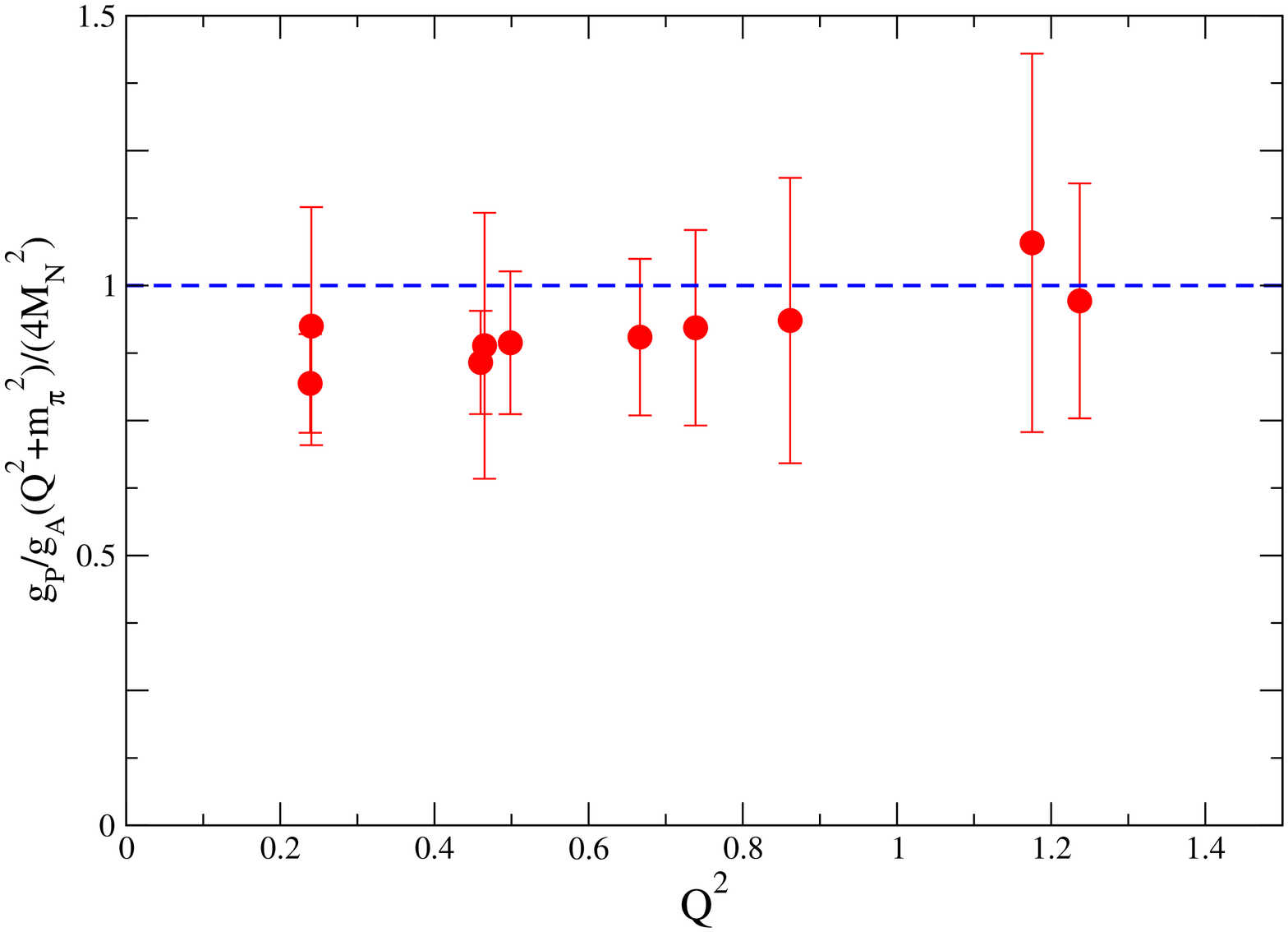}\vspace{0.35pc}
\caption{\label{GP_GA_pole} Isovector pion pole contribution, normalized as \hbox{ $(q^2+m_\pi^2)G_P / 4 m^2 G_A$}, at $m_\pi = $350 MeV .}
\end{minipage}
\end{figure}

The form factor $F_2$ is of particular interest following the
observation at JLab~\cite{Gayou:2001qd} that measurement of spin
transfer yields a form factor that decreases more slowly with
momentum transfer than the traditional Rosenbluth separation, which is
now believed to suffer from substantial contamination from two photon
exchange contributions. Figure~[\ref{F2_F1}] shows that the lattice
results indeed approach the experimental ratio $F_2/F_1$ from spin
transfer as the pion mass decreases.

The nucleon axial vector current form factors $G_A$ and $G_P$ are defined by
\begin{equation}
\langle p | \overline{\psi} \gamma^\mu \gamma_5 \psi | p' \rangle = \overline{u}(p)[ G_A(q^2)\gamma^\mu \gamma_5 + \frac{q^\mu}{2m} \gamma_5 G_P(q^2) + \sigma^{\mu\nu}\gamma_5 q_\nu G_M(q^2) ] u(p'). 
\end{equation}
Figure~[\ref{GA_F1}] shows lattice calculations of the isovector ratio
$G_A/F_1$ compared with experimental form factors extracted from pion
photoproduction and neutrino scattering.  Again, to within the roughly
10\% discrepancy between the two experiments, the lattice results are
qualitatively consistent with experiment. In the soft pion limit,
$G_P$ is dominated by the pion pole:
\begin{equation}
G_p(q^2) \sim \frac{4 m^2 G_A(q^2)}{q^2 +m^2_\pi}.
\label{pole}
\end{equation}
Figure~[\ref{GP_GA}] shows lattice results for the the ratio $G_P/G_A$
compared with experiment.  To demonstrate the degree to which
Eq.~[\ref{pole}] is satisfied, Fig.~[\ref{GP_GA_pole}] shows the ratio
$(q^2+m_\pi^2)G_P / 4 m^2 G_A$ for lattice data at pion mass 350 MeV,
which is consistent with unity.

\section{Generalized Parton Distributions}

\begin{figure}[tb]
\begin{center}
\includegraphics[width=17pc,angle=90,scale=1.8]{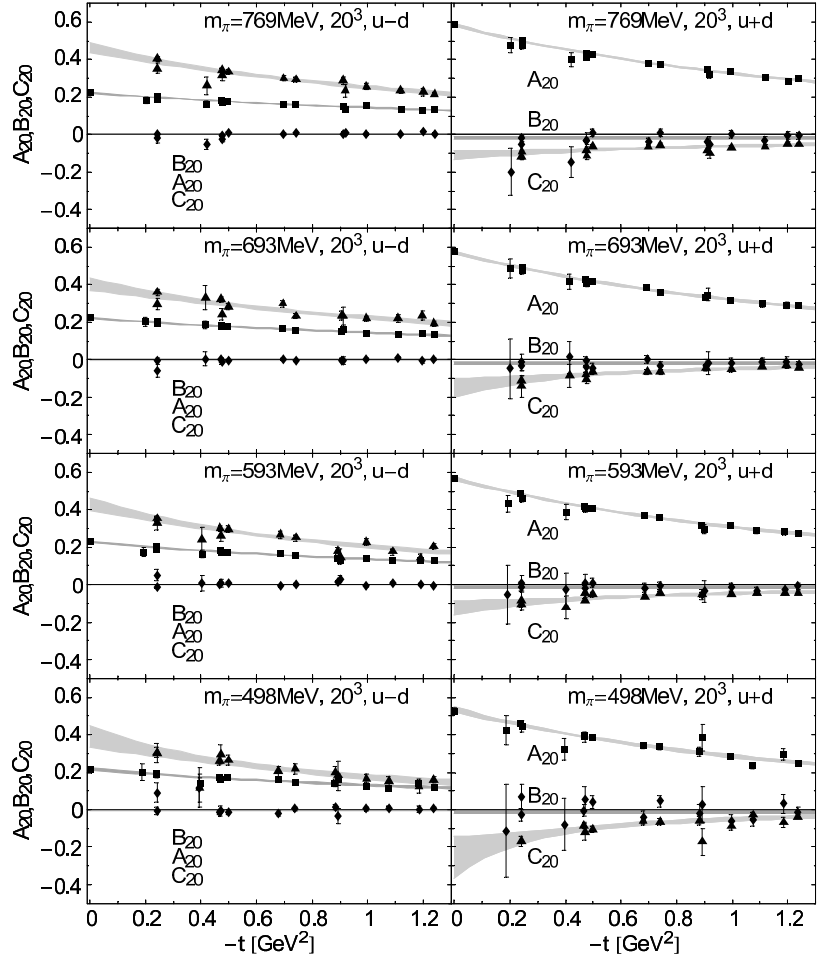}\vspace{1.8pc}
\caption{\label{ABC} Generalized form factors $A_{20}, B_{20}$ and $C_{20}$ for $u-d$ and $u+d$ at four masses.}
\end{center}
\end{figure}
Off diagonal matrix elements of the tower of operators in
Eq.~[\ref{LocalOp}] yield the generalized form factors
\begin{eqnarray}
 \langle P' | {\cal O}^{\mu_1} | P \rangle
 &=&\dlangle \gamma^{\mu_1 }\drangle A_{10}(t)
  + \frac{\imag}{2 m} \dlangle \sigma^{\mu_1
    \alpha} \drangle
  \Delta_{\alpha} B_{10}(t)\,, \nonumber \\ [.5cm]
 \langle P' | {\cal O}^{\lbrace \mu_1 \mu_2\rbrace} | P \rangle
 &=& \overline{P}{}^{\lbrace\mu_1}\dlangle
  \gamma^{\mu_2\rbrace}\drangle A_{20}(t)
  + \frac{\imag}{2 m} \overline{P}{}^{\lbrace\mu_1} \dlangle
  \sigma^{\mu_2\rbrace\alpha}\drangle \Delta_{\alpha} B_{20}(t)
  +\frac{1}{m}\Delta^{\{ \mu_1}   \Delta^{ \mu_2 \} }
  C_{20}(t)\,, \nonumber \\[.5cm]
  \langle P' | {\cal O}^{\lbrace\mu_1 \mu_2 \mu_3\rbrace} | P \rangle
  &=& \overline{P}{}^{\lbrace\mu_1}\overline{P}{}^{\mu_2} \dlangle
  \gamma^{\mu_3\rbrace}
  \drangle A_{30}(t)
  + \frac{\imag}{2 m} \overline{P}{}^{\lbrace \mu_1}\overline{P}{}^{\mu_2}
  \dlangle \sigma^{\mu_3\rbrace\alpha} \drangle
  \Delta_{\alpha} B_{30}(t)
  \nonumber \\
  &+& \Delta^{\lbrace \mu_1}\Delta^{\mu_2} \dlangle
  \gamma^{\mu_3\rbrace}\drangle A_{32}(t)
  + \frac{\imag}{2 m} \Delta^{\lbrace\mu_1}\Delta^{\mu_2}
  \dlangle \sigma^{\mu_3\rbrace\alpha}\drangle
  \Delta_{\alpha} B_{32}(t),
  \label{Para1}
\end{eqnarray}
where we use the short-hand notation $\dlangle \Gamma
\drangle=\overline U(P',\Lambda ')\Gamma U(P,\Lambda)$ for matrix
elements of Dirac spinors $U$ and where $\Delta=P'-P$ and
$t=\Delta^2$.

Of particular interest in this work is the relationship between the
generalized form factors and the origin of the nucleon spin.  The
contribution of the spin of the up and down quarks to the total spin
of the nucleon is given by the zeroth moment of the spin dependent
structure function $\langle 1\rangle_{\Delta q}$ as
\begin{equation}
\frac{1}{2} \Delta \Sigma = \frac{1}{2}\langle 1\rangle_{\Delta u + \Delta d}\,.
\label{spin}
\end{equation}
Note both that our previous calculation of $g_A$ confirms our ability
to calculate the connected contributions to $\langle 1\rangle_{\Delta
q}$ accurately on the lattice and that $\Delta \Sigma$ requires the
calculation of disconnected as well as connected contributions.
Throughout this section, we will only discuss the results of connected
diagrams, so all results will eventually need to be corrected for the
effect of disconnected diagrams as well.

\begin{figure}[tb]
\begin{minipage}{17.5pc}
\includegraphics[width=17pc,angle=0,scale=1]{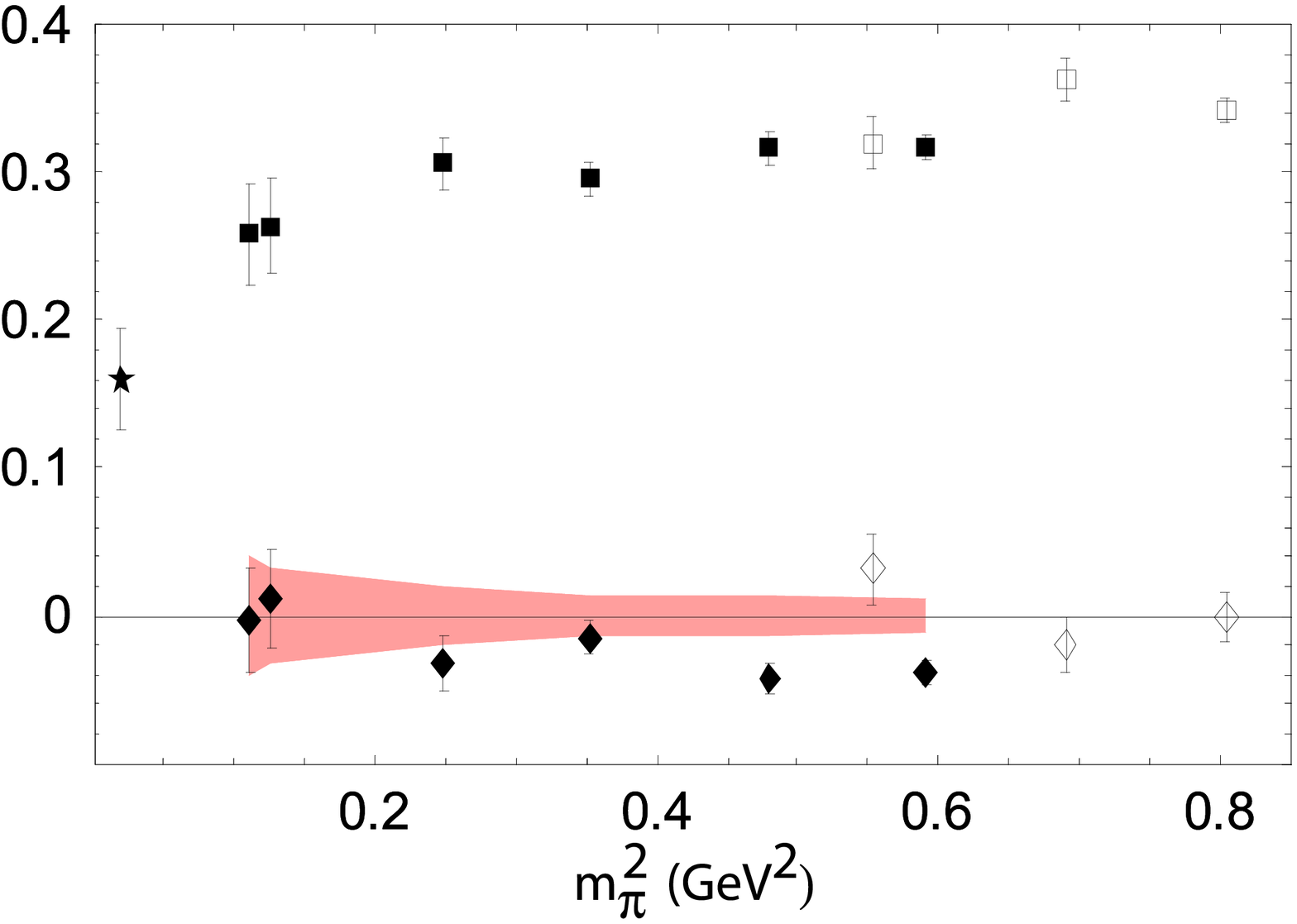}
\caption{\label{OAM1}Nucleon spin decomposition.  Squares denote $\Delta \Sigma^{u+d} /2$, the star indicates the experimental quark spin contribution, and  diamonds denote $L^{u+d}$. }
\end{minipage}
\hspace{0.5pc}
\begin{minipage}{17.5pc}
\includegraphics[width=17pc,angle=0,scale=1]{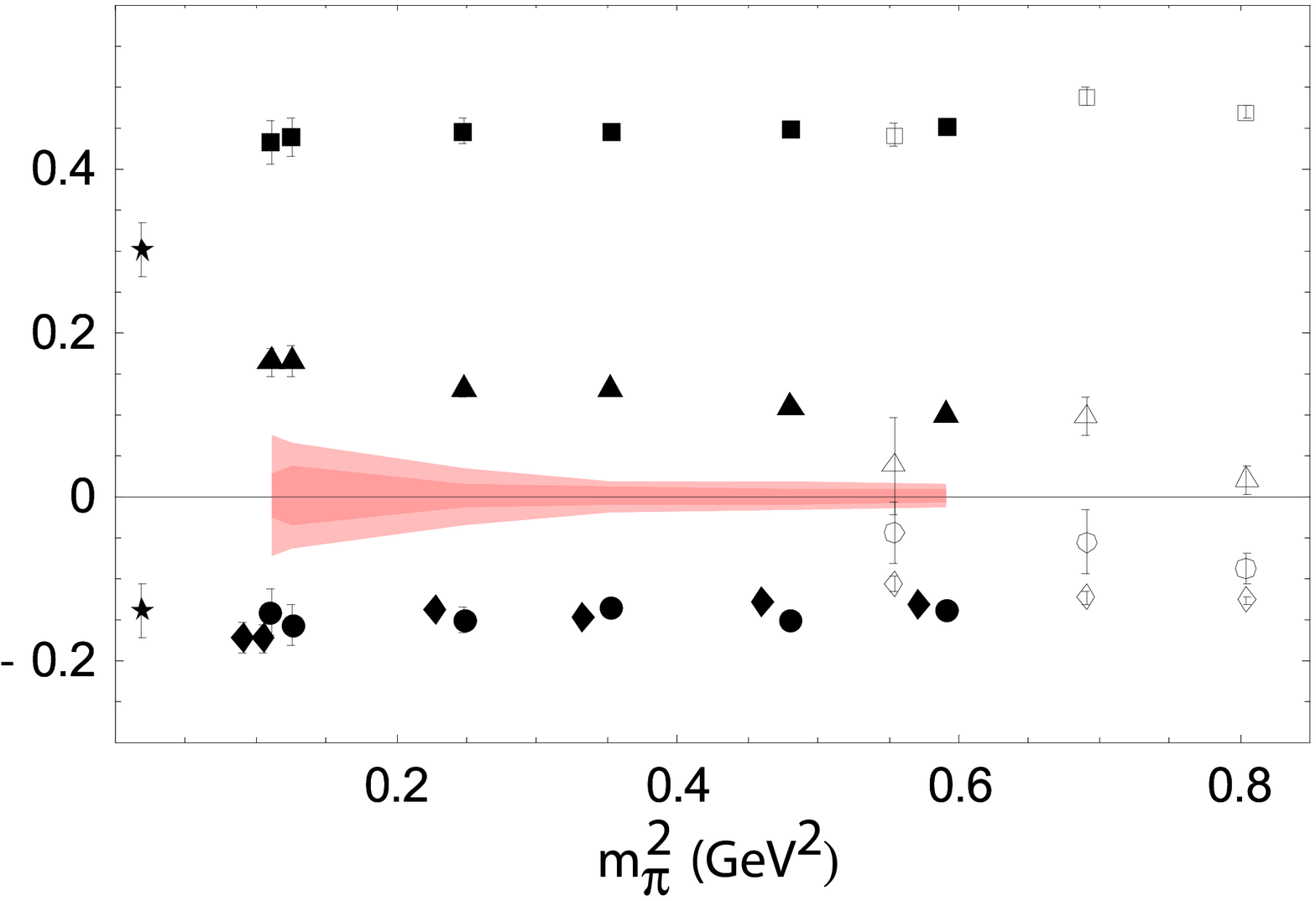}
\caption{\label{OAM2} Nucleon spin decomposition by flavor.  Squares denote $\Delta \Sigma^u /2$, diamonds denote $\Delta \Sigma^d /2$, triangles denote $L^u$, and  circles denote $L^d$.} 
\end{minipage}
\end{figure}

The total contribution to the nucleon spin from both the spin and
orbital angular momentum of quarks is given by the Ji sum
rule~\cite{Ji:1996ek}:
\begin{equation}
J_q = \frac{1}{2}\left(A_{20}^{u+d}(0) + B_{20}^{u+d}(0)\right),
\end{equation}
so that the contribution of the orbital angular momentum is given by
$L_q = J_q - \frac{1}{2}\Delta \Sigma$.  Earlier calculations in the
heavy quark regime showed that for 700 MeV pions, roughly 68\% of the
spin of the nucleon arises from the spin of quarks, 0\% arises from
orbital angular momentum, and hence the remaining 32\% must come from
gluons~\cite{Hagler:2003jd,Gockeler:2003jf,Mathur:1999uf}.  Figure~[\ref{ABC}] shows the recent
lattice data for $A_{20}$, $B_{20}$ and $C_{20}$ for lighter pion
masses, and the results for $\Delta \Sigma$ and $L_q $ are shown in
Fig.~[\ref{OAM1}].  The full decomposition showing the contribution of
spin and orbital angular momentum from up and down quarks is shown in
Fig.~[\ref{OAM2}].  Here, one observes that the angular momentum
contributions of both up and down quark are separately substantial,
and it is only the sum that is extremely small.

\begin{figure}[tb]
\begin{center}
\includegraphics[width=17pc,angle=0,scale=1.8]{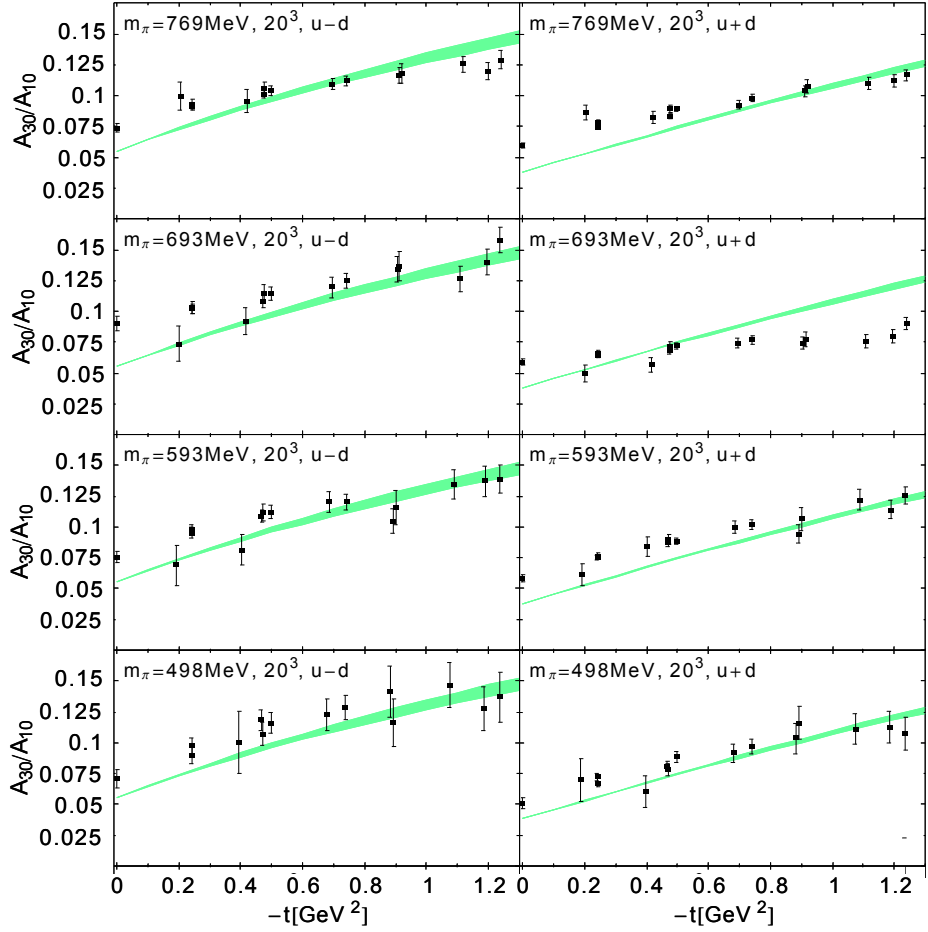}\vspace{2pc}
\caption{\label{A30_A10} Comparison of the ratio $ A_{30} / A_{10}$ for $u-d$ and $u+d$ at four masses with a phenomenological fit to generalized parton distributions. }
\end{center}
\end{figure}

In previous work in the heavy quark regime~\cite{Hagler:2003is}, we
have emphasized the dramatic differences in the slopes of $A_{10}$,
$A_{20}$, and $A_{30}$ and the fact that this reflects a sharp
decrease in the transverse size of the nucleon as $x$ approaches 1. To
show that this behavior also arises in the chiral regime,
Fig.~[\ref{A30_A10}] shows that the slope of the form factor ratio $
A_{30}/A_{10}$ differs substantially from unity as the pion mass is
decreased and also approaches the result given by a phenomenological
parameterization of the generalized parton
distributions~\cite{Diehl:2004cx}.

\section{Conclusions}

In summary, the hybrid combination of valence domain wall quarks on an
improved staggered sea has enabled us to begin to enter the era of
quantitative solution of full lattice QCD in the chiral regime.  The
axial charge, $g_A$, represents a successful, ``gold-plated'' test,
which argues well for the prospects of quantitative control of a range
of important nucleon observables. The chiral extrapolation of moments
of quark distributions using our self-consistently improved one-loop
analysis is encouraging, but of course we would still like to directly
calculate the turn over in the approach to the chiral regime.
Similarly, the general agreement between nucleon form factors of the
vector and axial currents is highly encouraging.  Generalized form
factors are also being calculated well into the chiral regime, and
give promise for understanding the origin of the nucleon spin and the
transverse structure of the nucleon.  In the long term, since lattice
calculations determine moments of generalized parton distributions and
experiments measure convolutions of generalized parton distributions,
there is an excellent opportunity for synergy between experiment and
theory in jointly determining generalized parton distributions. The
final analysis of all the data shown in this work is currently being
completed, and full results will be published in the near future.

This work also indicates obvious challenges for the future.  Clearly
the calculations must be extended to lower quark masses and finer
lattices and analyzed with partially quenched hybrid chiral
perturbation theory, and disconnected diagrams must be calculated.  In
addition, it is important to develop new techniques to explore form
factors at high momentum transfer, gluon observables, and transition
form factors for unstable states.
 
\section{Acknowledgments}

This work was supported by the DOE Office of Nuclear Physics under
contracts DE-FC02-94ER40818, DE-AC05-06OR23177 and DE-AC05-84150, the
EU Integrated Infrastructure Initiative Hadron Physics (I3HP) under
contract RII3-CT-2004-506078, the DFG under contract FOR 465
(Forschergruppe Gitter-Hadronen-Ph\"anomenologie) and the DFG
Emmy-Noether program.  Computations were performed on clusters at
Jefferson Laboratory and at ORNL using time awarded under the SciDAC
initiative. We are indebted to the members of the MILC collaboration
for providing the dynamical quark configurations which made our full
QCD calculations possible.

\bibliography{lat06.bib}

\end{document}